\documentclass[%
 reprint,
 amsmath,amssymb,
 aps,
 showkeys,
]{revtex4-2}

\usepackage{comment}
\usepackage[dvipdfmx]{graphicx}
\usepackage{dcolumn}
\usepackage{bm}
\usepackage{amsmath}
\usepackage{tikz-cd}
\usepackage[percent]{overpic}
\usepackage{multirow}
\usepackage{tikz}
\usetikzlibrary{arrows.meta, positioning, decorations.pathreplacing}
\usepackage{amssymb}

\usepackage{color}
\definecolor{Red}  {rgb}{1,0,0}
\definecolor{Green}{rgb}{0,1,0}
\definecolor{Blue} {rgb}{0,0,1}

\newcommand {\com}[1] {}

\usepackage{xcolor}
\usepackage{ulem}
\usepackage[draft, commandnameprefix=always]{changes}
\setaddedmarkup{{\color{red}#1}}
\setdeletedmarkup{\sout{#1}}

\begin{document}

\preprint{APS/123-QED}

\title{Group-Theoretic Upper Bounds on Reconstructability in Inverse Problems}

\author{Isshin Arai}%
\email{k078403@kansai-u.ac.jp}
\affiliation{1 Graduate School of Science and Engineering, Kansai University, Osaka, 564-8680, Japan}
\author{Tomoaki Itano}%
\affiliation{2 Department of Pure and Applied Physics, Faculty of Engineering Science, Kansai University, Osaka, 564-8680, Japan}

\date{\today}

\begin{abstract}
Reconstructing the causal structure of physical systems from observational data constitutes a fundamental inverse problem.
Here we show that the reconstruction dimension---defined as an upper bound on the number of recoverable components---is determined by the group-representation structure of the observation spaces and reconstruction maps.
This formulation provides an explicit and operational characterization of reconstructability and reconstruction dimension, extending ideas that are often understood only intuitively in equivariant representation theory.
As a concrete example, we demonstrate the reconstruction of the local velocity-gradient tensor from orientational measurements of particles suspended in flows, where the observation and velocity-gradient tensor spaces form SO(3) representations with constrained equivariant maps between them.
Using an SO(3)-equivariant neural network (implemented with e3nn), we show that the reconstructable subspaces predicted by the representation decomposition are qualitatively consistent with those found in practice.
Our formulation shows that the representation structure constrains reconstructability by determining an upper bound sector by sector, while our numerical results suggest that the actual saturation of the bound depends on the physics and data geometry.
Beyond providing a useful theoretical framework, this work also connects the abstract representation-theoretic structure to concrete inverse reconstruction problems in fluid physics.

\end{abstract}


\maketitle

\section{Introduction}\label{sec:intro}
In physics, a fundamental class of inverse problems involves inferring the underlying causal structure of a system (e.g., independent component analysis~\cite{Hy01}, scattering problems~\cite{Co98}, and identifiability 
analysis in dynamical systems~\cite{Wi21, Hei25, Mac19}).  
Such problems can also be formulated as inverse mapping problems, in which the input is reconstructed from the output by explicitly specifying the forward map~\cite{Al09, Ga22}.

We consider a physical system in which observables $X$ and 
$Y$ are related through a quantity $O$, which  characterizes the underlying cause. 
The central object of this work is the \textit{reconstruction 
map} $\mathfrak{F}$, which maps observation pairs $(X, Y)$ 
to $O$. We refer to this class of problems as 
\textit{inverse reconstruction}.
As a concrete example, we consider the problem of reconstructing the velocity-gradient tensor from the orientation and its time derivative of tracer particles in fluid flows~\cite{Sa85,Gau98}(see also Refs.~\cite{Dy82, Arai24} for flow 
visualization using particle orientations). Throughout this paper, we develop the analysis of the inverse reconstruction based on group-representation theory using this fluid-mechanical example as the primary illustration.

When addressing inverse problems in physics, at least the following two questions arise:
\begin{itemize}
\item $Q_1$ (Identifiability): Can $O$ be uniquely determined?
\item $Q_2$ (Reconstructability): How many components of $O$ can be recovered?
\end{itemize}
$Q_1$ has been commonly referred to as \textit{identifiability}~\cite{Mac19,Wi21, Hei25}~and is satisfied when the reconstruction map is well defined from the perspective of inverse reconstruction.
In this work, we focus on $Q_2$, which we term \textit{reconstructability}, and define the \textit{reconstruction dimension} as an upper bound on the number of components that can be resolved from a reconstruction map. 
Stability and accuracy are further considerations that accompany both questions.

Physical systems are strongly constrained by symmetry. 
When considering how symmetry restricts reconstruction maps and determines what information can be reconstructed, tools such as equivariant maps, intertwiners, and irreducible representations are standard.
For example, neural-network-based estimators that exploit system symmetries have been widely reported for both forward~\cite{Fo23, Batz23} and inverse~\cite{Ce21, Ta25} problems (see also Ref.~\cite{Ge22}).  
Existing SO(3)-equivariant networks such as e3nn are constructed based on group-representation decomposition.
However, these have not been explicitly formulated as a quantitative index such as the reconstruction dimension.

In this work, we formulate inverse reconstruction under symmetry in terms of equivariant maps between representation spaces.  
Inverse reconstruction can be defined in a purely mathematical setting without reference to any specific physics. However, once group-equivariance is imposed on the reconstruction map, the map itself reflects a kind of physical law. The equivariance condition expresses the symmetry of that law, and only under this constraint does the reconstruction become a reconstruction of that physics. From this viewpoint, we derive the reconstruction dimension.

We show that the number of independently resolvable 
components in each irreducible sector $\lambda$ is 
bounded above by $d_F^{(\lambda)}$, determined by the 
equivariant map design. For SO(3) with $N$ observation 
pairs, this gives $\min(N,\, 2\ell+1)$ 
(Fig.~\ref{fig:flow-rec-dim}).
This provides an upper-bound framework applicable to the physical system possessing symmetry. 

We further illustrate this framework of \textit{reconstruction dimension} numerically using an SO(3)-equivariant neural network, showing that 
the reconstruction dimension provides a trend consistent with a meaningful upper bound on reconstructability across irreducible components.
Our numerical illustration thus suggests that the proposed framework provides a theoretical guideline for representation selection and network design in equivariant neural networks.

The remainder of this paper is organized as follows.  
Section~\ref{sec:theory} formulates inverse reconstruction under symmetry in physics within the framework of group-representation theory.
Section~\ref{sec:results}  presents dynamics of particle orientation as examples of symmetric systems and their corresponding representation structures.  
Section~\ref{sec:analysis} analyzes the reconstruction dimension in SO(3) systems.
In section~\ref{sec:numeric}, the reconstruction map is approximated using an SO(3)-equivariant neural network and we numerically examine how the reconstruction dimension acts as an upper bound.
Section~\ref{sec:phys-noise} discusses the physical interpretation and noise robustness of the reconstruction.
Finally, section~\ref{sec:conclusion} summarizes our findings and outlines possible future directions.

\section{Inverse reconstruction under representation theory}\label{sec:theory}
In this section, we formulate inverse reconstruction under symmetry in physics from the viewpoint of group-representation theory and introduce the reconstruction dimension.
Let $X$ and $Y$ denote the observed input and output spaces, respectively, and let there exist a forward map $f_O$ representing the causal structure of the system with quantity $O$ that characterizes the underlying cause:
\[
f_O : X \to Y, \qquad 
x \in X, \; y = f_O(x) \in Y.
\]
Here, $f_O$ represents a physical process such as temporal evolution, chemical reaction, scattering, or measurement.  
 In this context, the goal is to infer the cause that determines $f_O$.
The cause may refer to the map $f_O$ itself or its constituent $O$.  
In particular, \textit{inverse reconstruction} is defined as the problem of constructing a map from the input–output pair space $X\times Y$ to the space $O$.
We define such a map as the \textit{reconstruction map} $\mathfrak{F}$ as follows:
\[
\mathfrak{F} : X \times Y \to O, \qquad 
(x,y) \mapsto o, \quad o \in O.
\]

In physical systems, both the input–output pair and the cause relationships are often subject to symmetry.  
This means that the system as a whole is invariant under the action of a symmetry group $G$, which imposes constraints on the reconstruction map $\mathfrak{F}$.  
Specifically, for any $g \in G$, the following commutative diagram must hold:
\begin{center}
  \includegraphics[width=0.7\linewidth]{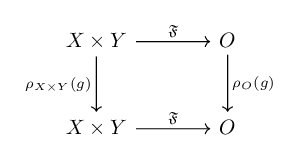}
\end{center}
where $\rho$ denotes a representation of the group $G$.  
Accordingly, an ideal reconstruction map $\mathfrak{F}$ satisfies
\[
\mathfrak{F} \circ \rho_{X\times Y}(g) = \rho_O(g) \circ \mathfrak{F}, 
\qquad \forall g \in G.
\]
Henceforth, $X$, $Y$, and $O$ are treated as representation spaces $V_X$, $V_Y$, and $V_O$ under the group action of $G$.  
In this work, we consider reconstruction maps of the form
\[
\mathfrak{F} = F_L \circ F,
\]
where $F$ is a nonlinear $G$-equivariant map
\[
F : V_X \times V_Y \to W,\label{eq:F}
\]
$W$ is an intermediate representation space, 
and $F_L \in \mathrm{Hom}_G(W, V_O)$ is a linear equivariant map 
(intertwiner), so that the composition $\mathfrak{F}$ is 
automatically equivariant.

Let $\widehat{G}$ denote the set of equivalence classes of 
irreducible representations of $G$, and for each 
$\lambda \in \widehat{G}$, let $W^{(\lambda)}$ and 
$V_O^{(\lambda)}$ denote the corresponding isotypic components 
in $W$ and $V_O$, respectively.
By Schur's lemma~\cite{Ful13}, the linear part $F_L$ acts as
\[
F_L : W = \bigoplus_{\lambda} W^{(\lambda)}
\;\longrightarrow\;
V_O = \bigoplus_{\lambda} V_O^{(\lambda)},
\]
block-diagonally with respect to the irreducible decomposition; that is,
\[
F_L\big|_{W^{(\lambda)}}
:\; W^{(\lambda)} \longrightarrow V_O^{(\lambda)},
\]
and no mixing between distinct irreducible components occurs.

We define the \textit{reconstruction dimension} associated with the map $F$ as follows.
For a fixed equivariant map $F$, the reconstruction dimension is
\begin{equation}
d_F^{(\lambda)} := \min\bigl(m_F(\lambda),\; \dim V_O^{(\lambda)}\bigr)
\in \mathbb{Z}_{\ge 0}, \quad \lambda \in \widehat{G},
\label{eq:eq-reconstrability}
\end{equation}
where $m_F(\lambda)$ is the multiplicity of $W^{(\lambda)}$ in the Clebsch--Gordan decomposition of $F$, determined solely by the representation structure of $F$.
The reconstruction dimension provides an upper bound on the number of resolvable components in each irreducible subspace.

\section{Application to particle orientation dynamics}\label{sec:results}

In this section, to illustrate the above framework, we consider a concrete system possessing SO(3) symmetry.
The SO(3) example is not essential to the theory; it simply provides a physically relevant demonstration of the group-theoretic reconstruction dimension framework. 

The orientational dynamics of a particle suspended in an incompressible fluid (in the limit where thermal fluctuations, shape effects, and inertia are negligible) is described as an SO(3)-symmetric system~\cite{Je1922, Got11, Arai25}:
\begin{equation}
\dot{\mathbf{s}} = \mathbf{s} \times \bigl(\mathbf{s} \times (\nabla \mathbf{u} \cdot \mathbf{s})\bigr),
\label{eq:orientation}
\end{equation}
where $\mathbf{s} \in \mathbb{S}^2$ denotes the unit normal vector of the  particle orientation, and $\nabla \mathbf{u}\in \mathcal{D}$ is the local velocity-gradient tensor, 
with $\mathcal{D}$ denoting the space of velocity-gradient tensors. 
The physical process map $f_{\nabla \mathbf{u}}$ can then be defined as
\begin{equation}
f_{\nabla \mathbf{u}} : \mathbb{S}^2 \to T_{\mathbf{s}}\mathbb{S}^2,  \qquad 
\mathbf{s} \mapsto \dot{\mathbf{s}}.
\label{eq:flake}
\end{equation}
For a finite set of observation directions $\mathcal{S}_N^* = \{{\mathbf{s}}_i\}_{i=1}^N$, the map is given by
\[
f^*_{\mathcal{S}_N^*} : \nabla \mathbf{u} \mapsto \{(\dot{\mathbf{s}}_i, \mathbf{s}_i)\}_{i=1}^N,
\qquad
\dot{\mathbf{s}}_i = \mathbf{s}_i \times \bigl(\mathbf{s}_i \times (\nabla \mathbf{u} \cdot \mathbf{s}_i)\bigr).
\]
The injectivity of $f^*_{\mathcal{S}_N^*}$ is guaranteed for $N \ge 4$ when the observation directions $\{\mathbf{s}_i\}$ are in general position, i.e., not confined to a single plane (thus including at least four non-coplanar directions).  
This addresses $Q_1$ (identifiability)\cite{Mac19}, whereas the reconstruction dimension addresses $Q_2$ (reconstructability).
Once the injectivity of $f^*_{\mathcal{S}_N^*}$ is ensured, the inverse map
\[
\mathfrak{F} := (f^*_{\mathcal{S}_N^*})^{-1} : \mathcal{C}_N \to \mathcal{D},
\]
is well defined, establishing a one-to-one correspondence between the set of observed pairs $\mathcal{C}_N$ whose elements are $\{(\dot{\mathbf{s}}_i, \mathbf{s}_i)\}_{i=1}^N$ and $\mathcal{D}$.  
The reconstruction map $\mathfrak{F}$ must also be equivariant under the group action:
\[
\mathfrak{F} \circ \rho_{\mathcal{C}_N}(R)
= \rho_{\mathcal{D}}(R) \circ \mathfrak{F},
\qquad \forall R \in \text{SO}(3).
\]

In what follows, we clarify the reconstruction dimension for the example introduced in section~\ref{sec:theory}, and then numerically show its trends by approximating the reconstruction map $\mathfrak{F}$ using an SO(3)-equivariant neural network, referred to as the \textit{Velocity Gradient Network} (VGN).

\subsection{Analysis of the reconstruction dimension}\label{sec:analysis}

The reconstruction map $\mathfrak{F}$ is constrained to the space of SO(3)-equivariant maps.
Observations are given as local observation pairs $(\mathbf{s}_i, \dot{\mathbf{s}}_i)$ for each particle.
We denote by $V_\ell$ the $(2\ell+1)$-dimensional vector space carrying the $\ell$th irreducible representation of SO(3).
In the example, $\mathbf{s}$ and $\dot{\mathbf{s}}$ transform as vectors (i.e., elements of $V_1$), while the velocity-gradient tensor $\nabla \mathbf{u}$ is a rank-2 Cartesian tensor transforming as $A \mapsto RAR^\top$, which belongs to the $V_1 \otimes V_1$ representation. Its decomposition $V_1 \otimes V_1 = V_0 \oplus V_1 \oplus V_2$ corresponds to the standard splitting into the trace (scalar), antisymmetric (vorticity), and symmetric-traceless (strain) components.
Furthermore, among SO(3)-equivariant maps, the tensor product $V_1 \otimes V_1$ provides a natural lowest-order polynomial construction.
We design the reconstruction map $\mathfrak{F}$ by 
composing the following three layers.\\

\paragraph{Layer 1: Bilinear map.}
For each observation pair $(\mathbf{s}_i, \dot{\mathbf{s}}_i) \in V_1 \times V_1$, we form the tensor product:
\begin{equation}
  (\mathbf{s}_i, \dot{\mathbf{s}}_i) \longmapsto \mathbf{s}_i \otimes \dot{\mathbf{s}}_i \in V_1 \otimes V_1.
  \label{eq:bilinear}
\end{equation}

\paragraph{Layer 2: Clebsch--Gordan decomposition.}
We decompose $V_1 \otimes V_1$ into irreducible representation spaces of SO(3):
\[
  V_1 \otimes V_1 = V_0 \oplus V_1 \oplus V_2.
\]
Projecting each observation pair yields
\begin{equation}
  \mathbf{s}_i \otimes \dot{\mathbf{s}}_i
  = a_i^{(0)} \oplus \mathbf{b}_i^{(1)} \oplus C_i^{(2)},
  \label{eq:CG_components}
\end{equation}
where the components correspond, respectively, to the scalar, antisymmetric, and symmetric-traceless parts under the SO(3) action. 
The composition of Layers 1 and 2 defines the nonlinear 
equivariant map $F: V_X \times V_Y \to W$, where the 
intermediate space $W = (V_0 \oplus V_1 \oplus V_2)^{\oplus N}$ 
is the direct sum of the Clebsch--Gordan decompositions over 
all $N$ observation pairs.

\paragraph{Layer 3: Weighted summation.}
The irreducible components from all $N$ pairs are combined 
by a weighted sum:
\begin{equation}
  {T}^{(\ell)} = \sum_{i=1}^{N} w_i^{(\ell)}\, 
  (V_\ell)_i,
  \qquad \ell = 0, 1, 2.
  \label{eq:weighted_sum}
\end{equation}
This corresponds to the linear equivariant map 
$F_L$, which acts on each 
isotypic component independently, as guaranteed by 
Schur's lemma.
Since each layer is SO(3)-equivariant, the full composition 
$\mathfrak{F} = F_L \circ F$ is also equivariant.\\

\begin{figure*}[t]
\centering
\includegraphics[width=0.9\linewidth]{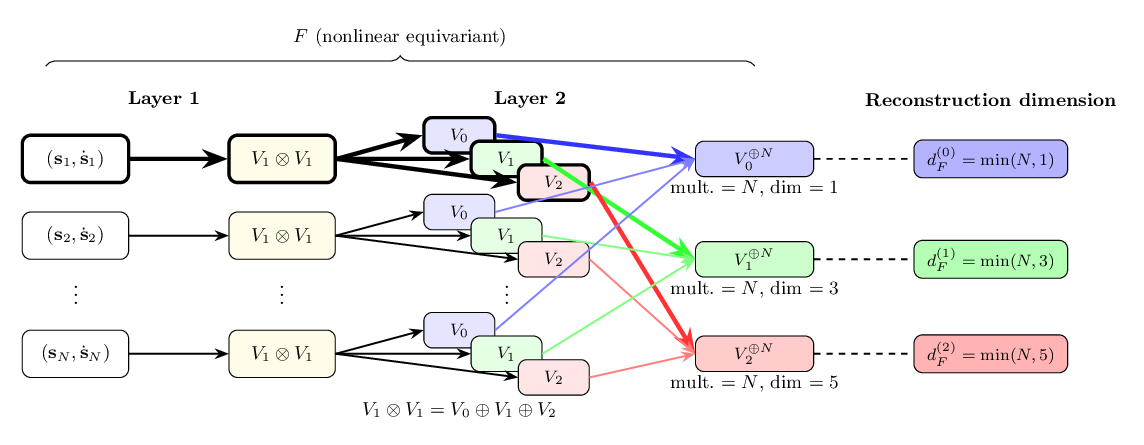}

\caption{Schematic flow of the reconstruction dimension with
$\mathfrak{F}=F_L\circ F$.
Each observation pair is decomposed into irreducible components
$V_0\oplus V_1\oplus V_2$ via the Clebsch--Gordan decomposition
(Layer~2) and projected
by the linear equivariant map $F_L$.
The reconstruction dimension
$d_F^{(\ell)}=\min(N,\,2\ell+1)$ reaches its maximum at
$N=1$, $3$, and $5$ for $V_0$, $V_1$, and $V_2$, respectively.
Bold lines highlight the first observation pair ($N=1$),
which already fills $V_0$ ($\dim V_0 = 1$) but provides
only one of the three components of $V_1$ and one of the five
components of $V_2$.}
\label{fig:flow-rec-dim}
\end{figure*}

The intermediate space $W$ 
has the same number of copies of each irreducible component 
$V_\ell$ as the number of observation pairs $N$,
while $V_O =V_0 \oplus V_1 \oplus V_2$ 
has dimension $2\ell + 1$ for each $V_\ell$.
As defined in~\eqref{eq:eq-reconstrability}, the reconstruction 
dimension $d_F^{(\ell)}$ for each irreducible component is 
therefore given by $\min(N,\, 2\ell + 1)$, as summarized 
in Fig.~\ref{fig:flow-rec-dim} and Table~\ref{tab:rec_dim}.
From Table~\ref{tab:rec_dim}, the reconstruction dimension  $d_F^{(1)}$ becomes the full dimension of $V_1$ corresponding to vorticity even for $N=3$, whereas the strain component $V_2$ remains partially unfilled. 
At $N=5$, the reconstruction dimension reaches the full dimension of three irreducible subspaces.
Thus, dimensional analysis based on representation structure provides the theoretical limit of information of inverse reconstruction.

By incorporating all pairwise combinations of observation directions, a tensor-product structure $(V_0 \oplus V_1 \oplus V_2)^{\otimes N}$ facilitates reaching the maximum most efficiently.
However, whether these modes are actually excited and appear linearly independent depends on the physical process and the observation configuration.
Hence, note that the fact that $d_F^{(\ell)}$
 reaches the full dimension of $V_{\ell}$ does not guarantee a good reconstruction.

\begin{table}[!htbp]
\caption{Theoretical reconstruction dimension $d_F^{(\ell)}$ for each representation component as a function of the number of observation pairs $N$ (for a map with a direct-sum structure).}
\label{tab:rec_dim}
\begin{ruledtabular}
\begin{tabular}{c|ccc|c}
$N$ & $V_0$ & $V_1$ & $V_2$ & Total \\ \hline
1 & 1 & 1 & 1 & 3 \\
2 & 1 & 2 & 2 & 5 \\
3 & 1 & 3 & 3 & 7 \\
4 & 1 & 3 & 4 & 8 \\
5 & 1 & 3 & 5 & 9 \\
\end{tabular}
\end{ruledtabular}
\end{table}

\subsection{Numerical Experiments}\label{sec:numeric}
In this section, we examine how the reconstruction dimension for each irreducible sector manifests itself in a neural-network implementation. Being an upper bound, it cannot be validated by the reconstruction error alone: a ground-truth tensor lying close to a reconstructable direction yields a small error irrespective of how many directions are actually spanned. Our aim here is therefore not to prove quantitative saturation, but to show, in a qualitative but consistent manner, that each irreducible sector independently respects its own upper bound.

\subsubsection{Dataset Generation and Simulation Setup}\label{subsub:setup}

The training and test datasets were numerically generated based on the orientational dynamics of particles suspended in an incompressible flow.  
Their motion is governed by Eq.~\eqref{eq:orientation}.
For each $\nabla \mathbf{u}$, Eq.~\eqref{eq:orientation} was directly integrated numerically with a second-order scheme and a time step of $\Delta t=2^{-14}$ to obtain the corresponding $\dot{\mathbf{s}}$.  
A total of $100$ orientation vectors $\{ \mathbf{s}_i \}_{i=1}^{100}$ were distributed nearly isotropically over the unit sphere, 
and for each velocity-gradient tensor, $N$ randomly selected pairs $\{(\dot{\mathbf{s}}_i, \mathbf{s}_i)\}_{i=1}^N$ were used as input samples.

Velocity-gradient tensors were generated from three representative spherical Couette flow (SCF) configurations: Axisymmetric flow, 2-fold spiral state, 3-fold spiral state~\cite{Go21, Arai24} (These flows are transitional states exhibiting distinct 
spatial patterns).
A total of $10^4$ samples were generated for each configuration, 
yielding $3\times 10^4$ velocity-gradient tensors in total.  
For each tensor, $100$ orientation samples were produced, resulting in $100\times 3\times 10^4$ orientation data in total.  
The dataset was split into 80\% for training and 20\% for validating.  
Additionally, test data from the 4-fold spiral state were used, comprising 24,500 velocity-gradient samples located within the same plane.

\subsubsection{Model Architecture and Training Conditions}\label{subsub:traininig}
\begin{figure}[!htbp]
    \centering
    \includegraphics[width=0.8\linewidth]{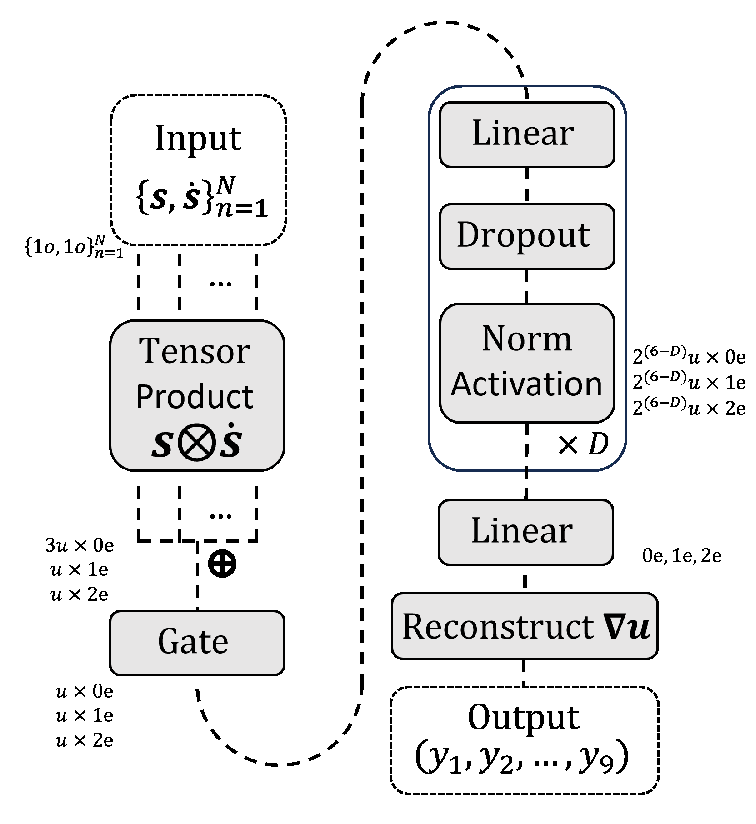}
    \caption{Schematic architecture of the Velocity Gradient Network (VGN).  
    The input consists of $N$ observation pairs 
$\{(\mathbf{s}_i, \dot{\mathbf{s}}_i)\}_{i=1}^N$, 
each represented as a pair of vector features 
($\{1o, 1o\}_{i=1}^N$ in e3nn notation).
For each pair $i$, a FullyConnectedTensorProduct produces 
tensor features 
($\{3u \times 0e,\, u \times 1e,\, u \times 2e\}_{i=1}^N$).
The $N$ pairs are then aggregated by a direct sum 
($\{3u \times 0e,\, u \times 1e,\, u \times 2e\}$), 
followed by a Gate layer that introduces nonlinearity 
through two series of $\tanh$ gates controlling the 
$\ell = 1$ and $\ell = 2$ tensor components 
($\{u \times 0e,\, u \times 1e,\, u \times 2e\}$).
Subsequently, $D$ layers of Linear, Dropout, and 
NormActivation are applied.
A final Linear layer maps the features to a 
nine-dimensional representation-space vector 
($\{0e,\, 1e,\, 2e\}$), from which the velocity-gradient 
tensor $\nabla \mathbf{u}$ is constructed; the $0e$ 
component is set to zero by the incompressibility 
constraint.\\
The simplified model VGN-DS is obtained by removing the 
Gate and the subsequent hidden layers, so that the 
direct-sum output is mapped to $\nabla \mathbf{u}$ by 
the final Linear layer alone.}
    \label{fig:arch}
\end{figure}

The Velocity Gradient Network (VGN)
was implemented using the e3nn library~\cite{Ge22} to ensure SO(3)-equivariance.  
Each $\mathbf{s}$ and $\dot{\mathbf{s}}$ is represented as a vector-type feature $V_1$ (denoted as {\it{1o}} in e3nn notation) 
and transformed through tensor products and gate mechanisms while preserving equivariance.
Figure~\ref{fig:arch} shows the schematic architecture of VGN.
The inclusion of cross terms via the FullyConnectedTensorProduct layer in e3nn increases the representational capacity of the network beyond the direct-sum structure assumed in the theoretical analysis.
Since the reconstruction dimension in Eq.~\eqref{eq:eq-reconstrability} is defined relative to a given map, the values in Table~\ref{tab:rec_dim} are the bounds for the direct-sum map $F$ consisting of Eqs.~\eqref{eq:bilinear} and \eqref{eq:CG_components} in section~\ref{sec:analysis} (Fig.~\ref{fig:flow-rec-dim}), not for every equivariant map. The cross terms place the VGN in a strictly larger map class, whose bound need not coincide with Table~\ref{tab:rec_dim}.
Nevertheless, the numerical results still capture the trend  predicted by the reconstruction dimension of Table~\ref{tab:rec_dim} (As shown later in Fig.~\ref{fig:model-comp} and Table~\ref{tab:RSE-comp}).
We employ the VGN, which is closer to the equivariant architectures used in practice, and whether the sector-by-sector behavior survives its cross terms is itself part of what we examine.

Hyperparameter optimization was performed under the condition $N=4$ by minimizing the mean squared error (MSE) 
using the Adam optimizer, 
where batch size $B$, depth $D$, and number of hidden units $u$ were varied sequentially, 
and the validation MSE was compared (Table~\ref{tab:hyper}) .  
Although increasing $u$ further reduced the loss, we fixed $u=128$ since the focus of this work is 
not on performance optimization but on observing the relationship between reconstruction dimension and the number of input pairs $N$.

\begin{table}[!htbp]
\caption{Hyperparameter search.}
\begin{ruledtabular}
\begin{tabular}{lcc}
Parameter & Search space & Final value \\ \hline
Batch size $B$ & 32, 64, 128 & 32 \\
Depth $D$ & 1, 2, 3 & 2 \\
Units $u$ & 16, 32, 64, 128, 256 & 128 \\
\end{tabular}
\end{ruledtabular}
\label{tab:hyper}
\end{table}

The evaluation metrics are defined as follows.
The normalized mean squared error (nMSE), 
relative squared error (RSE), 
mean relative squared error (MRSE), 
and covariance equivariance error (CovErr) are given by
\begin{align*}
\text{nMSE} &:=
\frac{\sum_j \|\hat{T}_j - T^{\text{true}}_j \|_F^2}{\sum_j \|T^{\text{true}}_j\|_F^2}, \\
\text{RSE}_j &:=
\frac{\|\hat{T}_j - T^{\text{true}}_j \|_F^2}{\|T^{\text{true}}_j\|_F^2}, \\
\text{MRSE} &:=
\frac{1}{M}\sum_j \text{RSE}_j, \\
\text{CovErr} &:=
\frac{1}{M} \sum_j \| \hat{T}_j^{\textrm{rot}} - R_j T^{\rm true}_j R_j^\top\|_F.
\end{align*}
Here, $M$ denotes the number of samples, $\hat{T}_j$ 
and $T^{\text{true}}_j$ denote the reconstructed and 
true tensors, respectively, and 
$\hat{T}_j^{\mathrm{rot}}$ is the tensor reconstructed 
from the rotated input 
$\{(R_j\mathbf{s}_i,\, R_j\dot{\mathbf{s}}_i)\}_{i=1}^N$, 
where $R_j$ is the rotation applied to the $j$th sample.
The Frobenius norm is defined as 
$\|A\|_F^2 = \sum_{i,j} A_{ij}^2$.

\subsubsection{Reconstruction Results}\label{subsub:results}

This section presents the reconstruction results for the 4-fold spiral state in SCF.  
For comparison, a standard multilayer perceptron (MLP) was also 
implemented and optimized over the same hyperparameter search space 
($B$, $D$, and $u$) 
under identical training conditions.
Each input pair $(\mathbf{s}_i, \dot{\mathbf{s}}_i)$ is concatenated and processed 
through several fully connected layers with tanh-activation and dropout 0.1 to output the reconstructed tensor.
\begin{table}[!htbp]
\caption{Comparison of reconstruction performance among models for $N=4$.}
\label{tab:comparison}
\begin{ruledtabular}
\begin{tabular}{lcc}
Metric & VGN & MLP\\ \hline
nMSE & $0.41$ & $0.48$ \\
MRSE & $0.41$ & $0.52$\\
CovErr & $2.54\times10^{-4}$ & $9.09\times10^{1}$ \\
\end{tabular}
\end{ruledtabular}
\end{table}
As shown in Table~\ref{tab:comparison}, compared with the MLP, which exhibits similar reconstruction errors (nMSE and MRSE),
the covariance error (CovErr) of the VGN is orders of magnitude smaller,  
demonstrating that the equivariant design stabilizes rotational variance in the reconstruction process.
Both the VGN and MLP were trained and validated under comparable conditions,
with hyperparameters adjusted to achieve stable convergence of the validation loss.
However, neither model was fully optimized for performance.

\begin{figure*}[t]
\centering
\begin{overpic}[width=1\columnwidth]{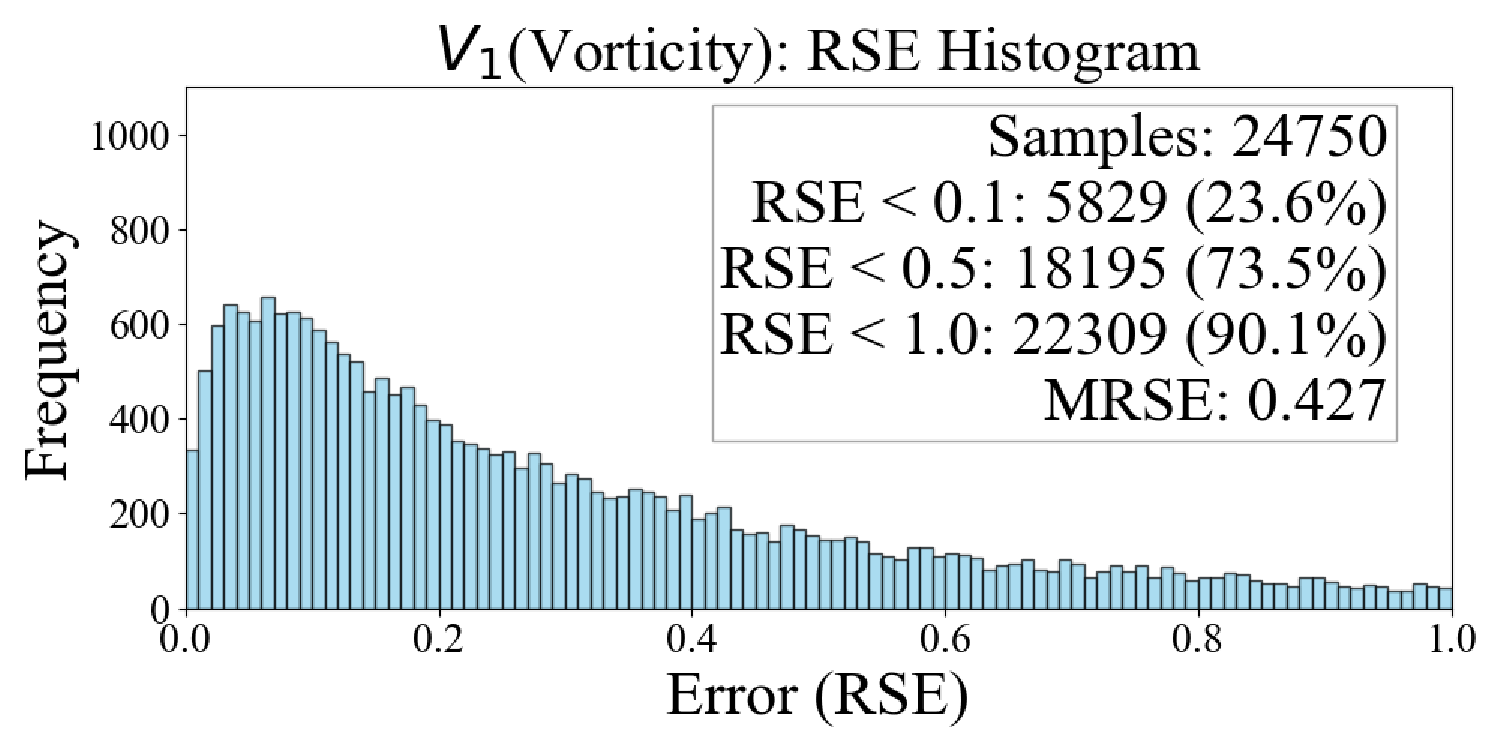}
  \put(5,48){\small (a)}
\end{overpic}
\hfill
\begin{overpic}[width=1\columnwidth]{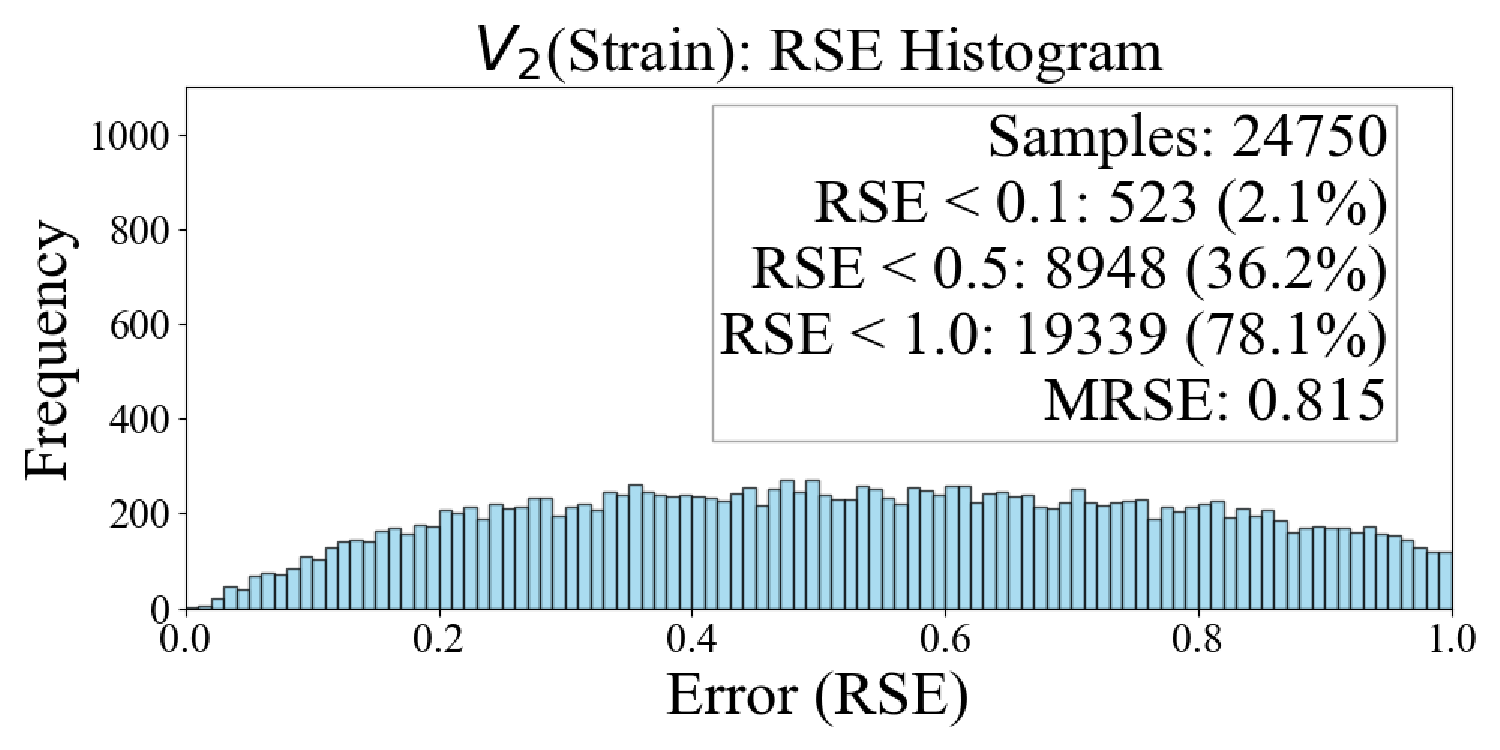}
  \put(5,48){\small (b)}
\end{overpic}
\caption{RSE histograms for $N=3$.
(a) Vorticity component $V_1$, (b) strain component $V_2$.
The $V_1$ component shows high reconstruction accuracy, whereas the $V_2$ component exhibits a broad error distribution.}
\label{fig:hist_N3}
\end{figure*}

\begin{figure*}[t]
\centering
\begin{overpic}[width=1\columnwidth]{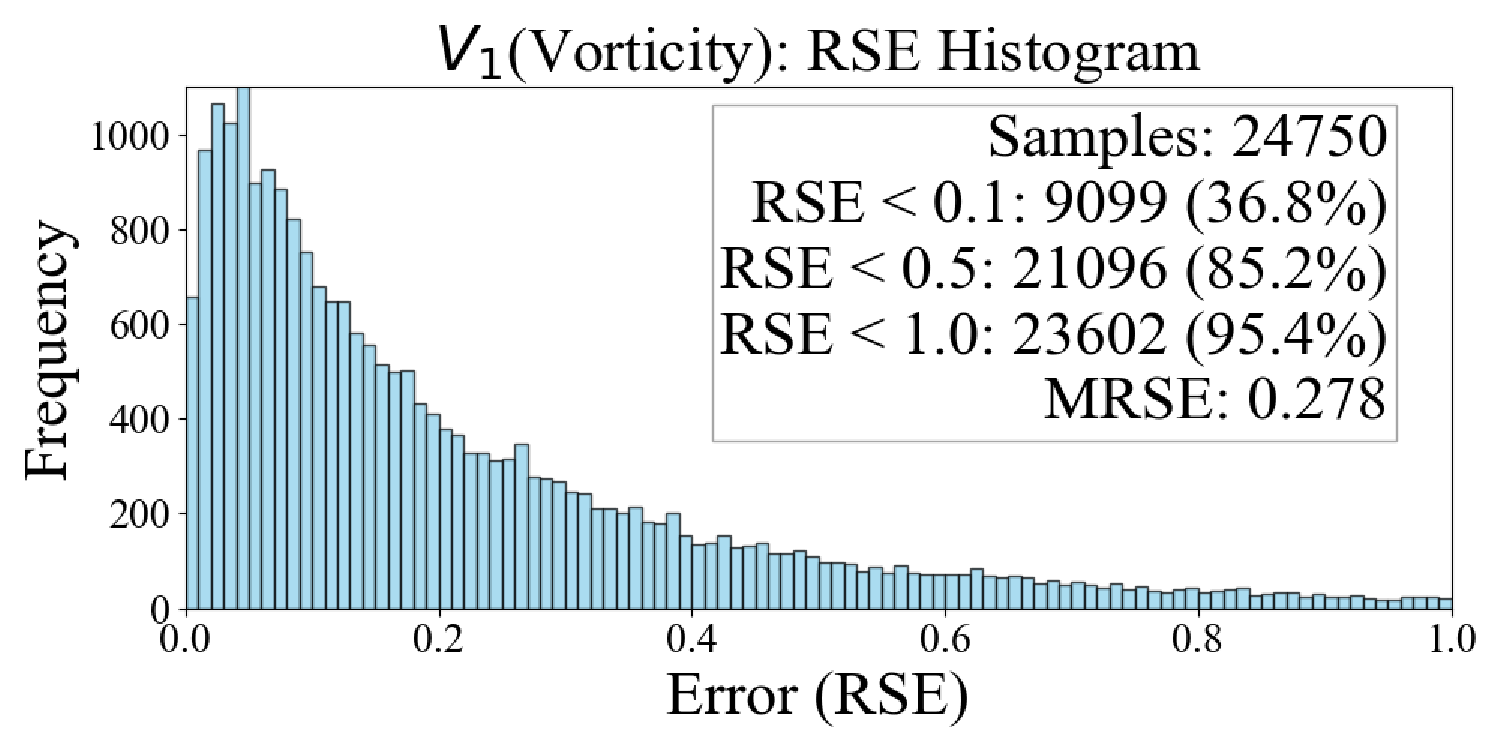}
  \put(5,48){\small (a)}
\end{overpic}
\hfill
\begin{overpic}[width=1\columnwidth]{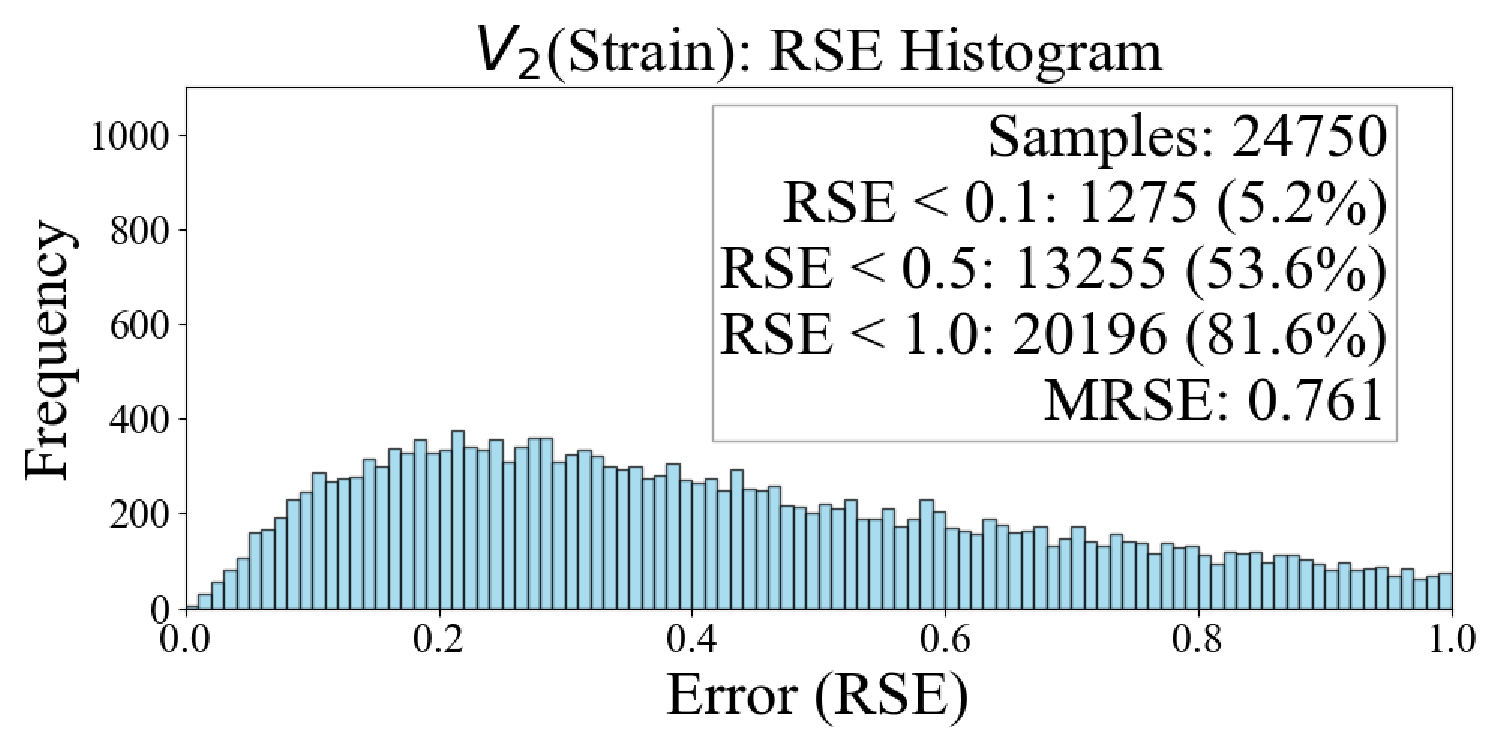}
  \put(5,48){\small (b)}
\end{overpic}
\caption{RSE histograms for $N=5$.
(a) Vorticity component $V_1$, (b) strain component $V_2$.
The distribution of $V_2$ shifts toward lower error, indicating partial reconstruction progress.}
\label{fig:hist_N5}
\end{figure*}

\begin{table}[!htbp] 
\caption{Reconstruction errors for $V_1$ (vorticity) and $V_2$ (strain) components.} \label{tab:comparison_N} 
\begin{ruledtabular} 
\begin{tabular}{c|cc|cc|cc} & \multicolumn{2}{c|}{$N=3$} & \multicolumn{2}{c|}{$N=4$} & \multicolumn{2}{c}{$N=5$}\\ Component & $V_1$ & $V_2$ & $V_1$ & $V_2$ & $V_1$ & $V_2$ \\ \hline nMSE & $0.27$ & $0.68$ & $0.21$ & $0.63$ & $0.17$ & $0.56$\\ MRSE & $0.43$ & $0.82$ & $0.31$ & $0.80$ & $0.28$ & $0.76$\\ 
\end{tabular} 
\end{ruledtabular} 
\end{table}

Table~\ref{tab:comparison_N} summarizes the reconstruction errors (nMSE, MRSE) 
for $N=3,4,5$ in each representation component.  
Figures~\ref{fig:hist_N3} and \ref{fig:hist_N5} show the corresponding RSE distributions.
As shown in Fig.~\ref{fig:hist_N3}, for $N=3$, 
the antisymmetric component $V_1$ (vorticity) is reconstructed accurately, 
while the symmetric traceless component $V_2$ (strain) exhibits a broad error distribution.  
In contrast, for $N=5$, in Fig.~\ref{fig:hist_N5}, both $V_1$ and $V_2$ distributions shift toward lower errors, 
indicating partial reconstruction of $V_2$ as $N$ increases.  
These trends are consistent with the theoretical reconstruction dimension in Table~\ref{tab:rec_dim}, 
supporting that the reconstruction dimension functions as an upper bound on the number of independently resolvable components in each irreducible subspace.
We emphasize that the error values themselves are not small (Table~\ref{tab:comparison_N}). These results should be understood as a qualitative trend in which each sector independently respects its upper bound, rather than as a demonstration that the bound is quantitatively saturated. The MLP ($N=4$) shows no such separation between sectors, with an nMSE of about $0.49$ for $V_1$ and $0.47$ for $V_2$. Since the MLP is not equivariant, its output does not decompose into the irreducible sectors, and the reconstruction dimension derived from the equivariant map $\mathfrak{F}$ does not apply to it.

\begin{figure}[htbp]
\centering
\includegraphics[width=1\columnwidth]{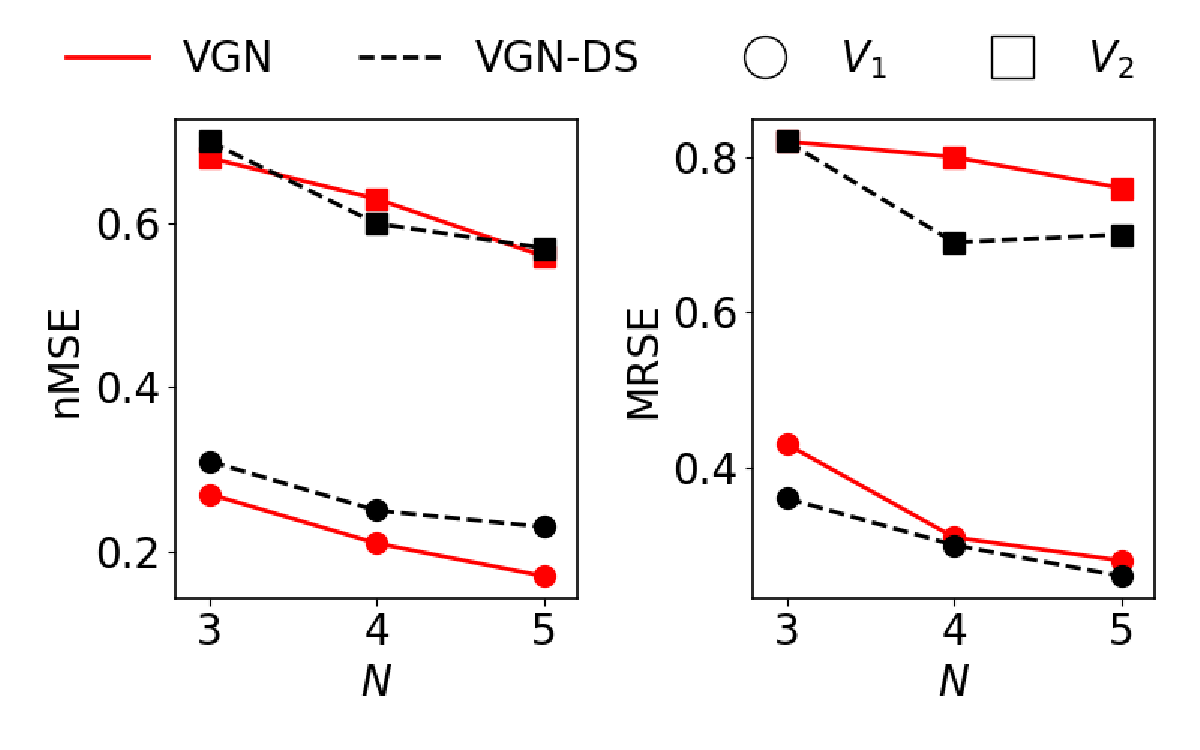}
\caption{Comparison of nMSE and MRSE between VGN (solid lines) and 
VGN-DS (dashed lines) for $N = 3, 4, 5$. Circle and square 
markers correspond to the $V_1$ and $V_2$ components, 
respectively.
For each $N$ and each component, the nMSE is lower for VGN, 
while the MRSE is lower for VGN-DS.
However, when evaluated at the level of individual samples, 
the fraction of samples achieving RSE $< 0.1$ is consistently 
higher for VGN (Table~\ref{tab:RSE-comp}), indicating that 
the hidden layers improve reconstruction accuracy for 
well-reconstructed samples.}\label{fig:model-comp}
\end{figure}

To isolate the contribution of the cross terms in the VGN 
architecture (Fig.~\ref{fig:arch}), we also implemented a 
simplified model, VGN-DS, in which the Gate and deep nonlinear 
layers are removed while retaining only the direct-sum structure.
The parameter $B$ and $u$ were kept identical to those 
of the VGN.
Figure~\ref{fig:model-comp} and Table~\ref{tab:RSE-comp} show 
that VGN and VGN-DS exhibit similar overall error levels, 
with VGN achieving moderately better performance in terms 
of nMSE (Fig.~\ref{fig:model-comp}) and the fraction of low-RSE samples (Table~\ref{tab:RSE-comp}).
Crucially, both models display the same qualitative pattern: 
$V_1$ is reconstructed with higher accuracy 
than $V_2$, and both components improve as $N$ increases 
from 3 to 5.
This suggests that the reconstruction dimension derived
by the direct-sum analysis (Table~\ref{tab:rec_dim}) 
constrains the reconstruction behavior regardless of whether 
cross terms are present.

\begin{table}[!htbp]
\caption{Fraction (\%) of samples with RSE below $0.1$ for $N=5$.}
\label{tab:RSE-comp}
\begin{ruledtabular}
\begin{tabular}{c|cc}
Component & VGN & VGN-DS \\ \hline
$V_1$ & 36.76 & 25.08 \\
$V_2$ & 5.15  & 4.01  \\
\end{tabular}
\end{ruledtabular}
\end{table}

The equivariant structure between representation spaces determines the reconstruction dimension as an upper bound, while the physical process $f_O$ and observation configuration may further restrict the effectively accessible subspace, reducing the achievable reconstruction dimension below this upper bound.
We further discuss in section~\ref{sec:phys-noise} that the reconstruction behavior depends on the underlying physics, data geometry, and noise.

\section{Physical interpretation and robustness}\label{sec:phys-noise}
\begin{figure}[!htbp]
\centering
\includegraphics[width=0.9\columnwidth]{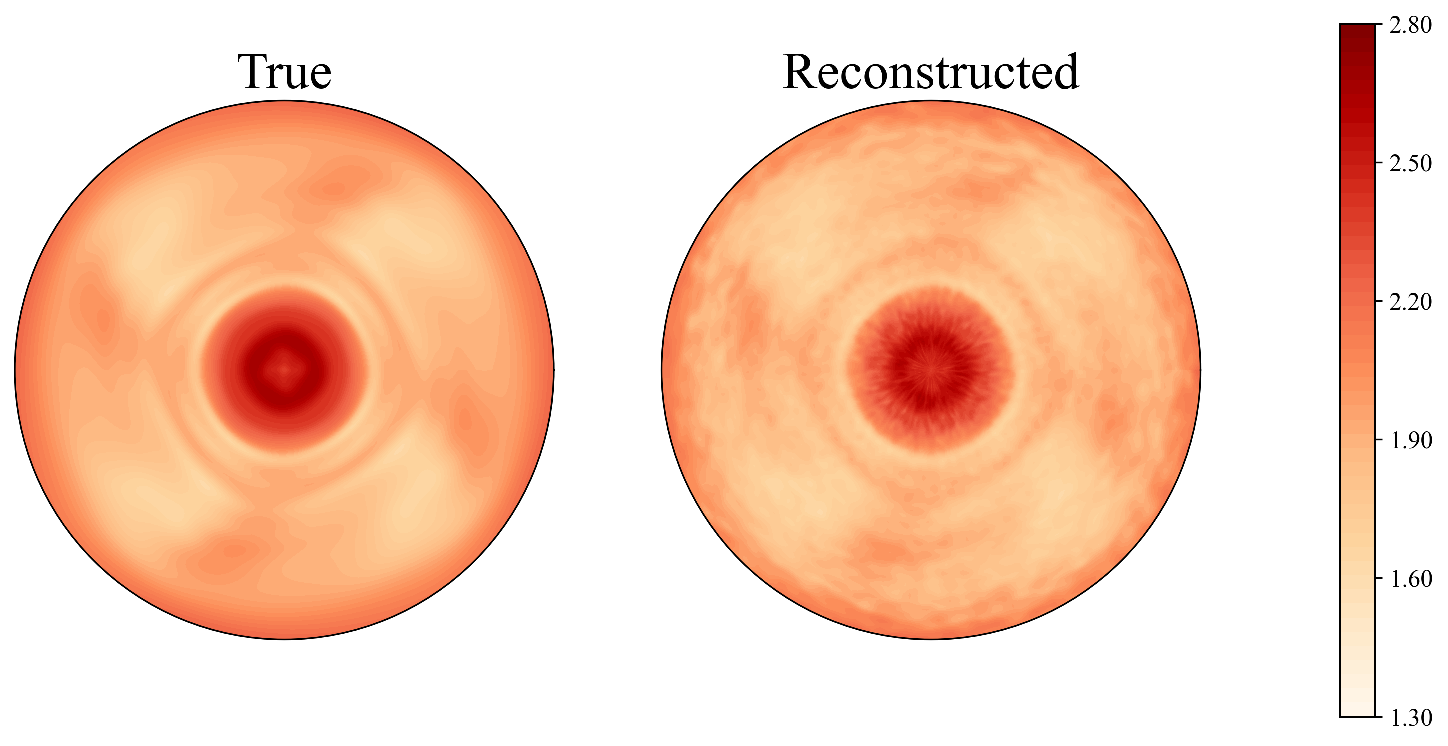}
\includegraphics[width=0.9\columnwidth]{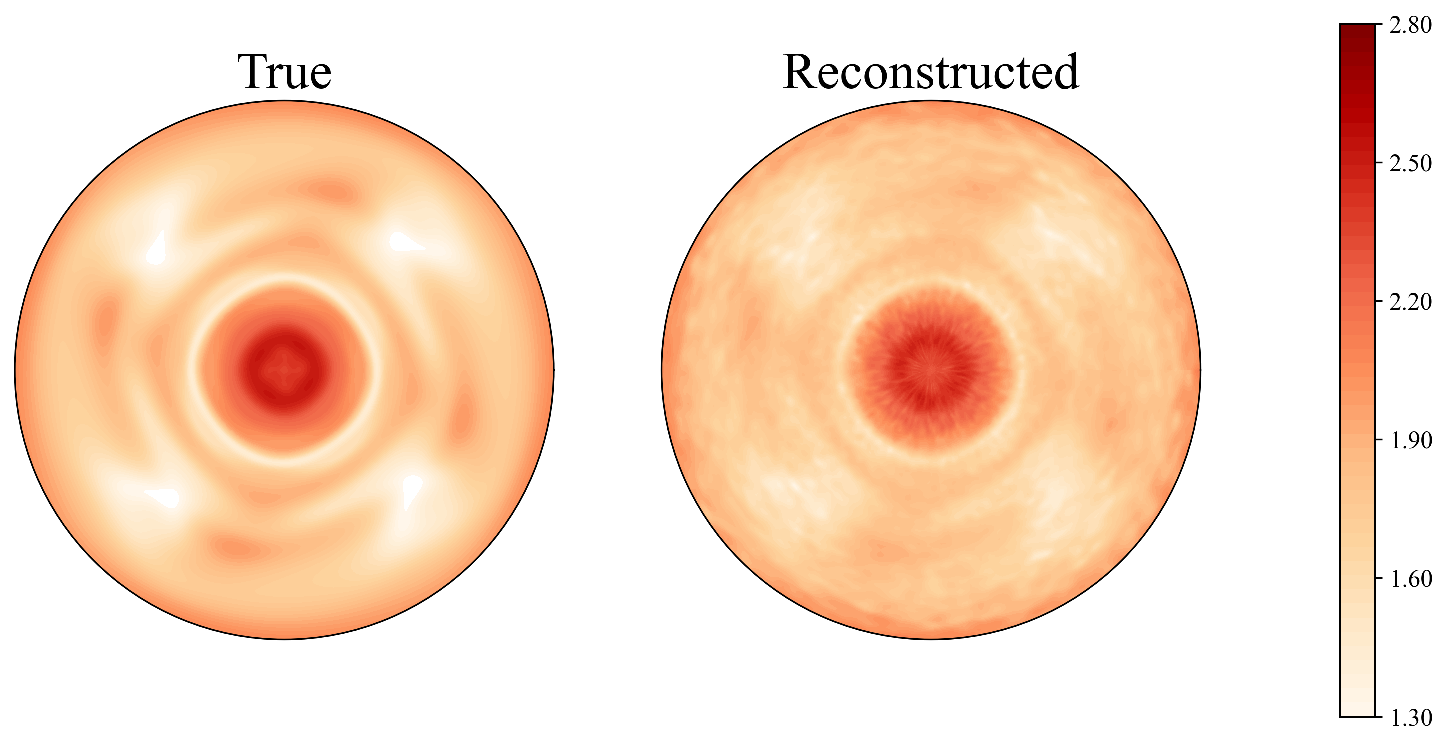}
\includegraphics[width=0.9\columnwidth]{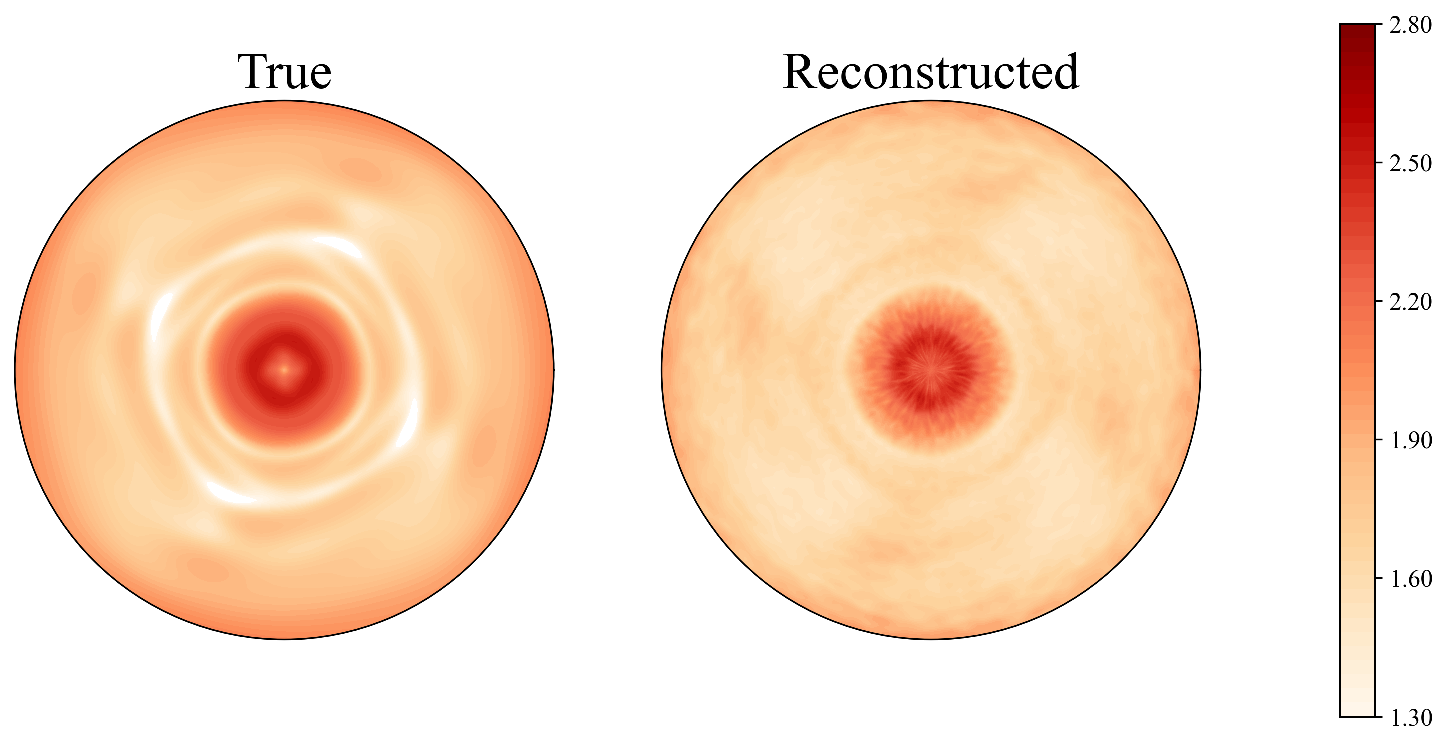}
\caption{
Comparison of Frobenius-norm ($\log_{10}$-scaled) spatial distributions of velocity-gradient tensors and their components ($N=5$).  
(a) Total tensor, (b) vorticity component $V_1$, and (c) strain component $V_2$.  
Left: ground truth; right: VGN reconstruction.}
\label{fig:velocitygrad_components}
\end{figure}
Figure~\ref{fig:velocitygrad_components} shows spatial distributions of the Frobenius-norm of reconstructed velocity-gradient tensors for the $N=5$ case in the 4-fold spiral state in SCF:  
(a) the total velocity-gradient tensor, (b) vorticity component $V_1$, and (c) strain component $V_2$, 
with the left and right panels showing the ground truth and VGN reconstructions, respectively.
The antisymmetric component $V_1$ (vorticity) exhibits a spatial pattern similar to the total tensor norm, 
whereas the symmetric component $V_2$ (strain) shows distinct spatial variation.  
This indicates that the dominant spatial mode of this dataset originates from the rotational component $V_1$. The analyzed system corresponds to particle orientation dynamics in SCF, where the inner sphere rotates while the outer sphere is stationary. 
At the level of the particle 
dynamics, the physical mapping Eq.~\eqref{eq:orientation} strongly depends on 
the $V_1$ component, as it enters the driven phase equation 
as the natural angular velocity~\cite{Arai26}. 
This illustrates that reconstructability is not determined by the reconstruction dimension alone. Although the $V_2$ component attains its full reconstruction dimension at $N=5$ (Table~\ref{tab:rec_dim}), its reconstruction error remains markedly large, reflecting the dominance of $V_1$ as the physically excited hydrodynamic mode.

\begin{figure}[!htbp]
\centering
\begin{overpic}[width=0.9\columnwidth]{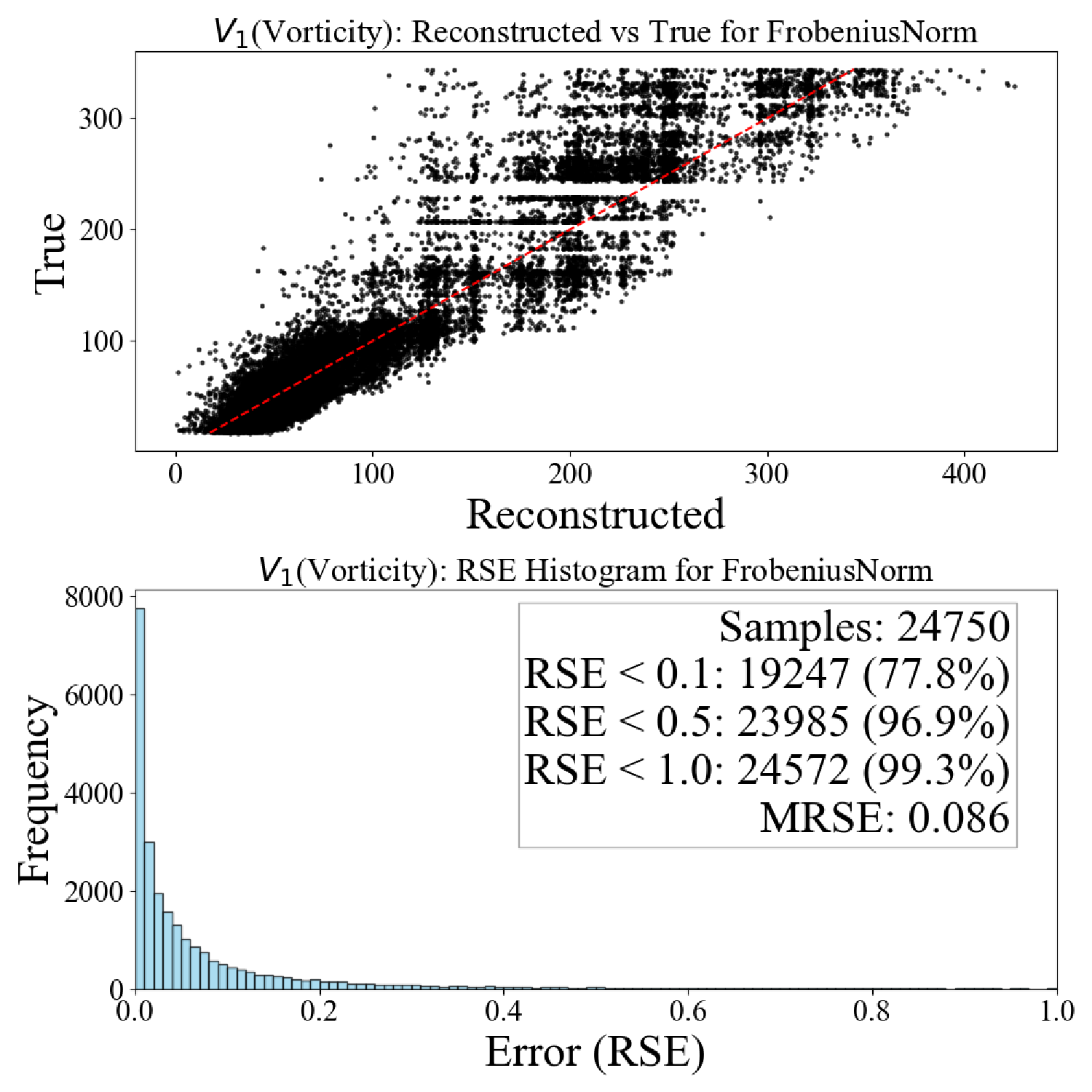}
  \put(5,100){\large (a)}
\end{overpic}

\vspace{10mm} 

\begin{overpic}[width=0.9\columnwidth]{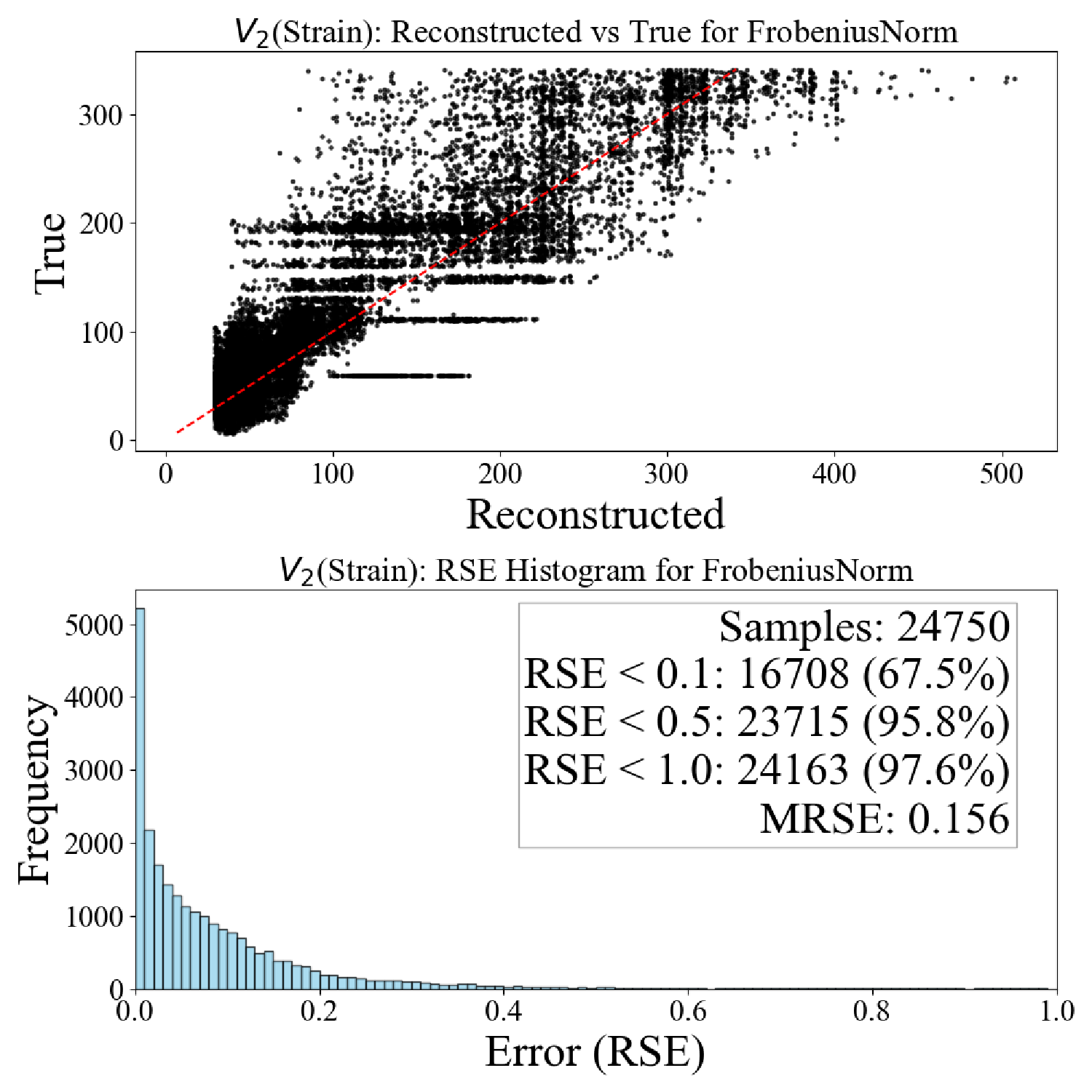}
  \put(5,100){\large (b)}
\end{overpic}
\caption{
Relationship between the true and predicted Frobenius norms of the velocity-gradient tensor components, and their RSE distributions ($N=5$).
(a) Vorticity component $V_1$, (b) strain component $V_2$.
Top panels show the correlation between true and predicted values, and bottom panels show the RSE histograms.
}
\label{fig:Frobenius Norm_N_5}
\end{figure}
Figure~\ref{fig:Frobenius Norm_N_5} shows the correlations between $\|\hat{T}_i\|_F$ and $\|T_i^{\mathrm{true}}\|_F$, and corresponding RSE distributions,
for the $V_1$ and $V_2$.
Spatial variation is dominated by $V_1$, whose reconstruction accuracy is higher, 
while $V_2$ also shows improving agreement with the true values when evaluated in terms of $\|T^{\text{true}}_i\|_F$.
\begin{figure*}[t]
\centering
\begin{overpic}[width=1\columnwidth]{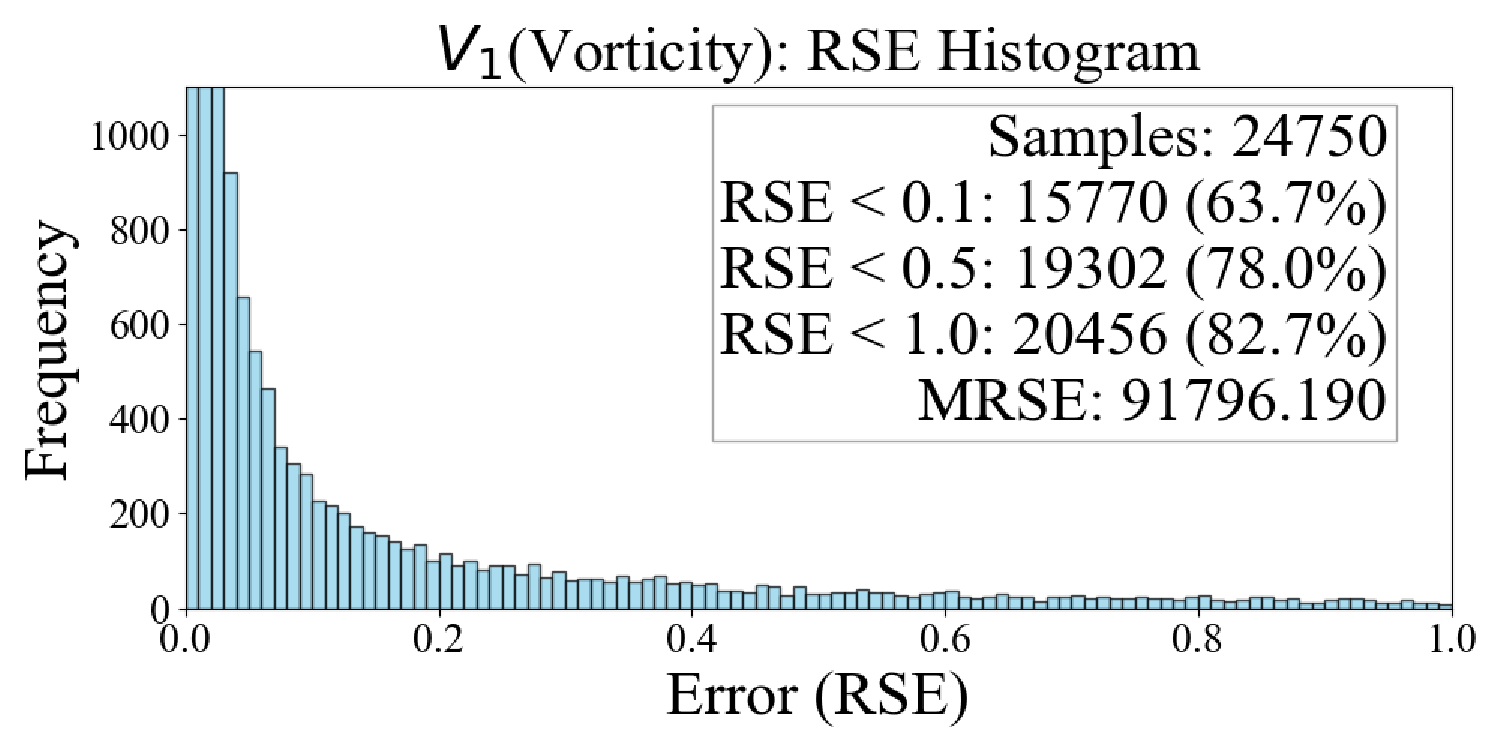}
  \put(5,48){\small (a) $\sigma=10^{-4}$}
\end{overpic}
\hfill
\begin{overpic}[width=1\columnwidth]{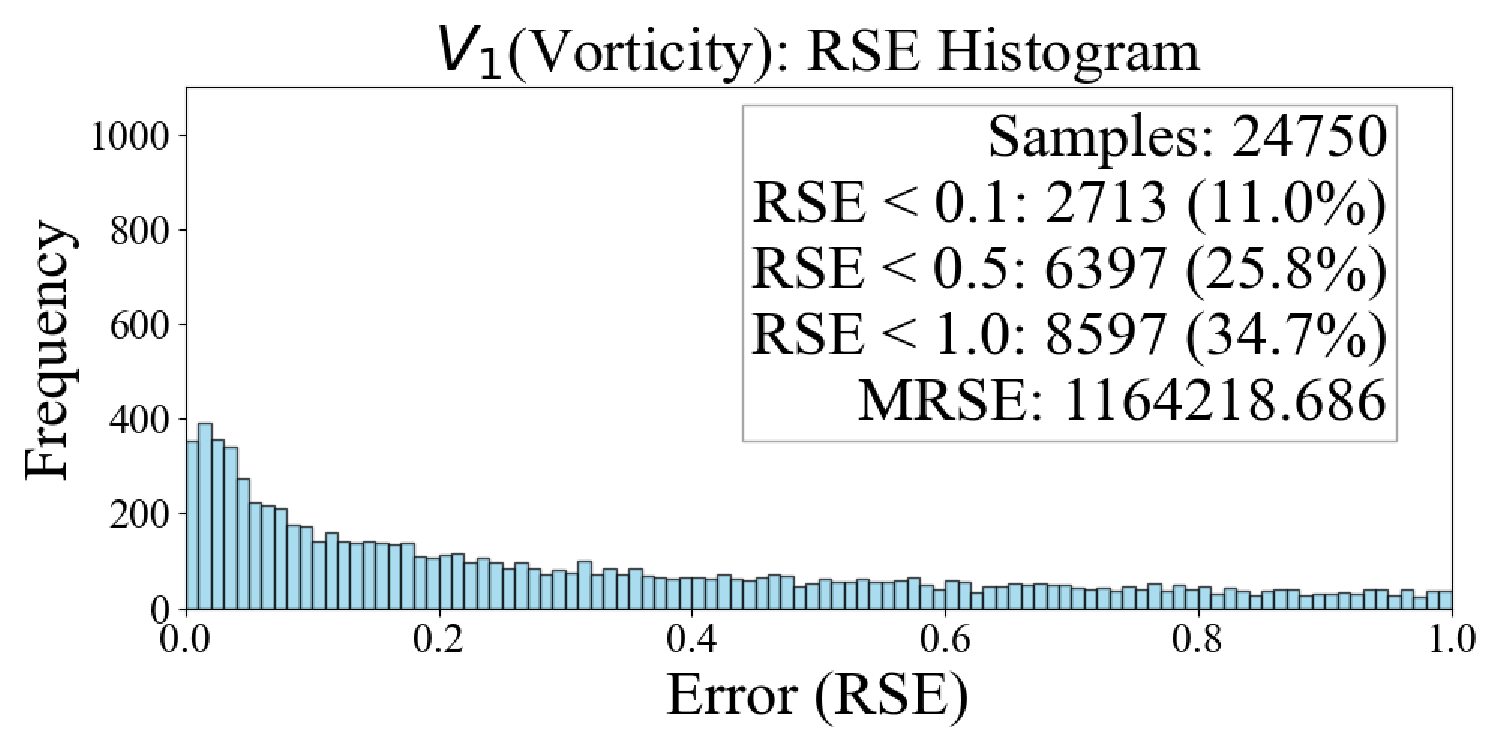}
  \put(5,48){\small (b) $\sigma=10^{-3}$}
\end{overpic}

\caption{
RSE histograms for the analytic reconstruction under Gaussian noise.
(a) $\sigma=10^{-4}$, (b) $\sigma=10^{-3}$.
The error distribution broadens as the noise level increases.
}
\label{fig:hist_analyze_noise}
\end{figure*}

\begin{figure*}[t]
\centering

\begin{overpic}[width=1\columnwidth]{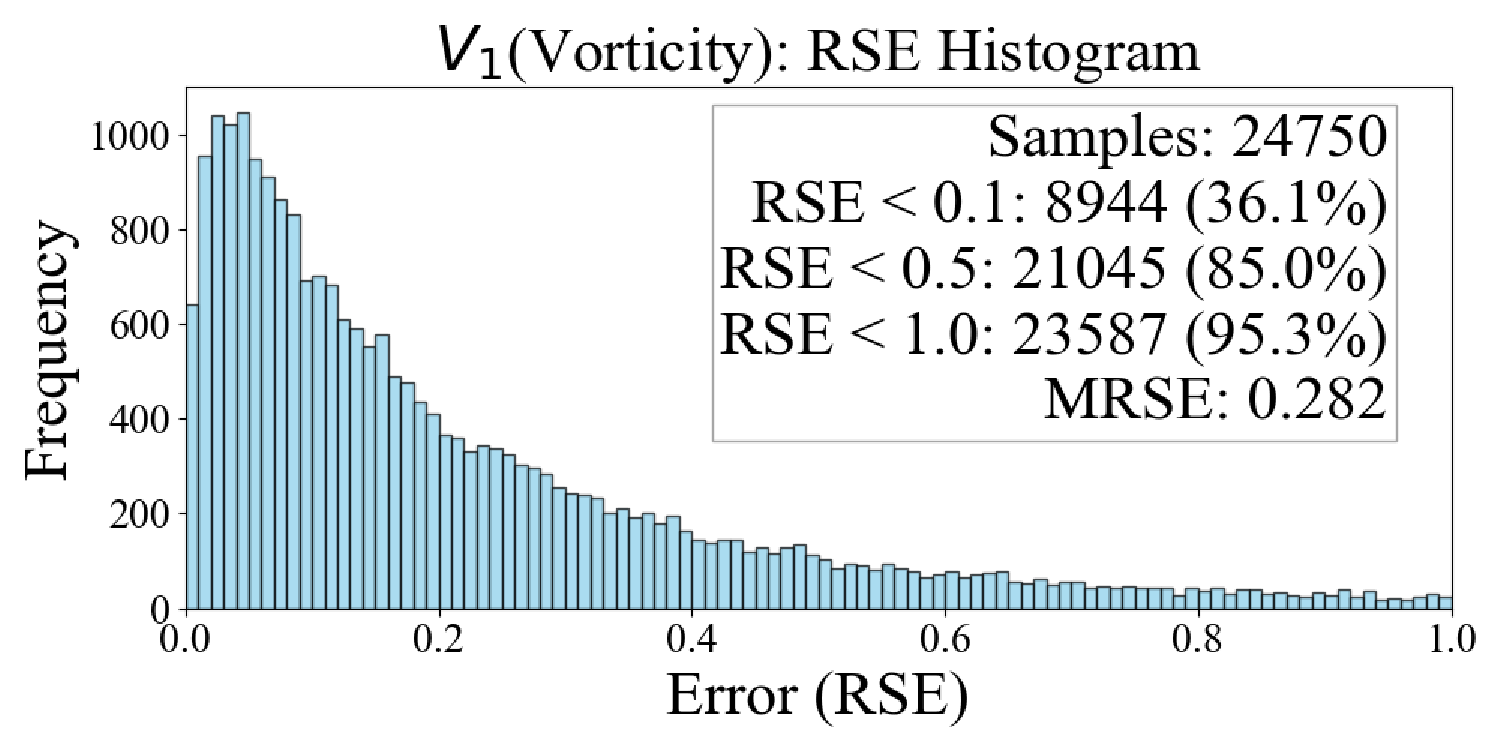}
  \put(5,48){\small (a) $\sigma=10^{-4}$}
\end{overpic}
\hfill
\begin{overpic}[width=1\columnwidth]{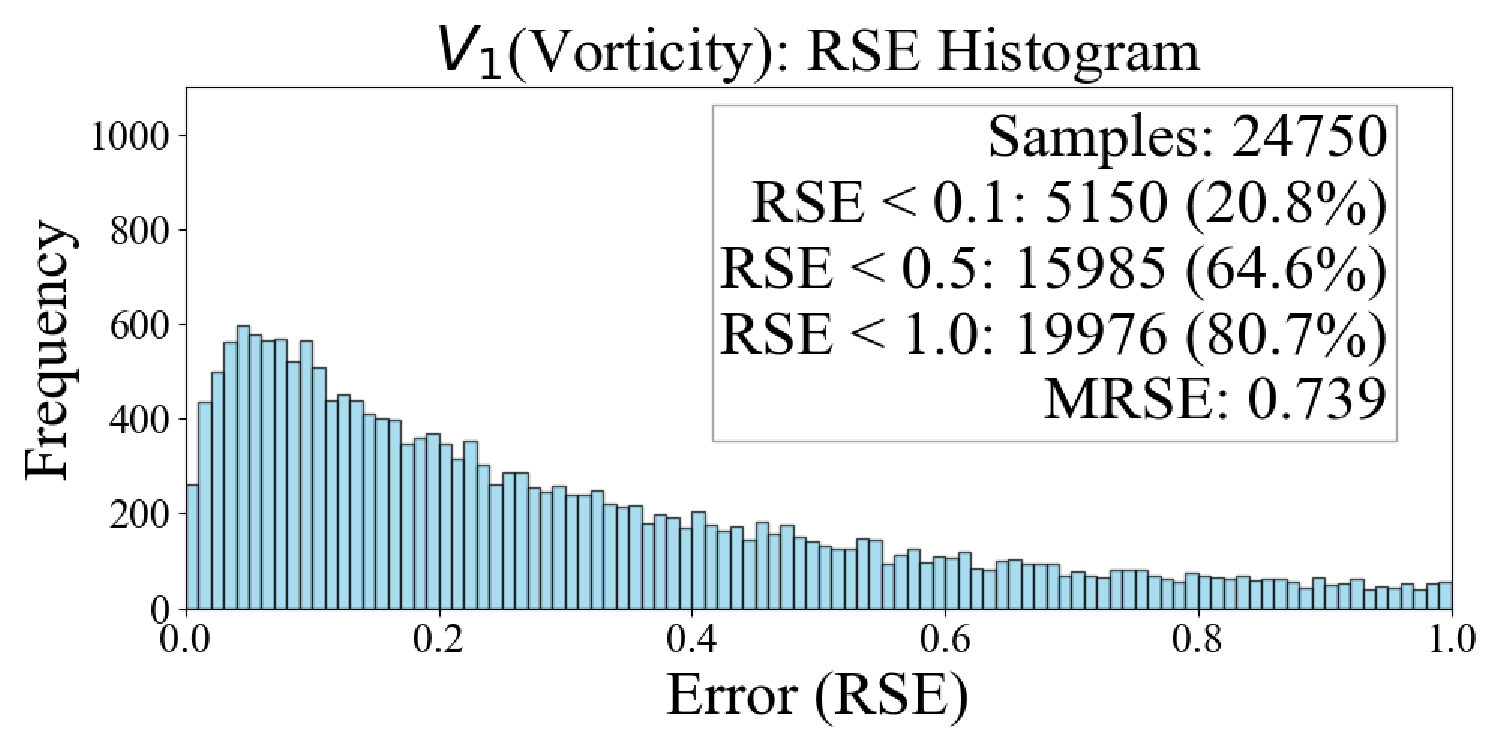}
  \put(5,48){\small (b) $\sigma=10^{-3}$}
\end{overpic}

\caption{
RSE histograms for the VGN reconstruction ($N=5$) under Gaussian noise.
(a) $\sigma=10^{-4}$, (b) $\sigma=10^{-3}$.
High reconstruction accuracy is maintained even under noisy conditions.
}
\label{fig:hist_vgn_noise}
\end{figure*}
Finally, Gaussian noise with mean 0 and standard deviation $\sigma$ was added to the orientation vectors $\mathbf{s}_i$, 
and reconstruction performance of the $V_1$ component was evaluated.  
The resulting RSE distributions are shown in Figs.~\ref{fig:hist_analyze_noise} and~\ref{fig:hist_vgn_noise}.
Figure~\ref{fig:hist_analyze_noise}(a),(b) shows the noise dependence in analytic reconstruction.  
For $\sigma = 10^{-3}$, the ratio of samples with RSE $<1$ drops to $35.2\%$, indicating high sensitivity to noise.  
In contrast, as shown in Fig.~\ref{fig:hist_vgn_noise}(a),(b), 
the VGN with $N=5$ maintains $80.7\%$ of samples with RSE $<1$ under the same noise condition, 
demonstrating that the equivariant mapping of VGN provides strong robustness to input noise.

\section{Conclusion}\label{sec:conclusion} 

In this study, we formulated the inverse problem in physical systems with symmetry as inverse reconstruction within the framework of group-representation theory.  
We showed that the symmetry imposes an equivariance constraint on the reconstruction map $\mathfrak{F}$, and that this constraint determines the reconstruction dimension $d_F^{(\lambda)}$ (as defined in~\eqref{eq:eq-reconstrability})---an upper bound on the number of recoverable components---through the decomposition structure of the representations, via Schur's lemma.

As a concrete example, we considered the orientational dynamics of advected particles in a fluid---an SO(3)-symmetric system---and constructed the mapping $\mathfrak{F} : \mathcal{C}_N \to \mathcal{D}$ between the observation and velocity-gradient tensor spaces.  
Using an SO(3)-equivariant neural network, the Velocity Gradient Network (VGN), we implemented the reconstruction of the local velocity-gradient tensor and illustrated that the theoretically derived reconstruction dimension was reflected as a qualitative trend during the network training process.
We showed that the representation structure constrains reconstructability by determining an upper bound for each sector, while the actual saturation of the bound depends on the physics and data geometry.

We provided a concrete and operational formulation of ideas in equivariant representation theory that are often understood only intuitively.
Future work includes extending this framework to higher-order or non-compact group structures, exploring systems with different symmetries, and formulating the degeneracy of the reconstruction dimension due to physical processes.

\begin{acknowledgments}
The authors would like to thank Mr. F. Goto and Mr. T. Inagaki for their pilot surveys.
We gratefully acknowledge discussions with Prof. Masako Sugihara-Seki.
This work was supported in part by a Grant-in-Aid for Scientific Research(C) and JSPS KAKENHI Grant No.24K07331 and by the Kansai University Grant-in-Aid for progress of research in graduate course, 2024.
\end{acknowledgments}

\bibliography{group_inverse}

\end{document}